\documentclass[10pt,journal]{IEEEtran}
\usepackage{textcomp}
\usepackage[pdftex]{graphicx}
\graphicspath{{../}}
\DeclareGraphicsExtensions{.pdf,.jpeg,.png,.eps}
\usepackage[table,xcdraw]{xcolor}
\usepackage{fixltx2e}
\usepackage[english]{babel}
\usepackage[utf8]{inputenc}
\usepackage{algorithm}
\usepackage[noend]{algpseudocode}
\usepackage{titlesec}

\usepackage{subfloat}
\usepackage{epstopdf}
\usepackage{cite}
\usepackage[normalem]{ulem}
\usepackage{caption}
\usepackage{subcaption}
\usepackage[cmex10]{amsmath}
\usepackage{float,multirow}
\usepackage{amsfonts}
\usepackage{comment}
\usepackage{framed}
\usepackage[export]{adjustbox}

\usepackage[euler]{textgreek}
\usepackage{epsfig,bm,amsmath,amssymb,graphicx,setspace,amsfonts}
\usepackage{algorithm}
\usepackage{color}
\usepackage[utf8]{inputenc}
\usepackage[T1]{fontenc}
\usepackage{algpseudocode}
\usepackage{textgreek}
\usepackage{amsthm}
\usepackage{array}
\usepackage{balance}
\usepackage{booktabs,subcaption,amsfonts,dcolumn}
\usepackage{enumerate}
\usepackage{listings}
\usepackage{epstopdf}
\def\BibTeX{{\rm B\kern-.05em{\sc i\kern-.025em b}\kern-.08em
		T\kern-.1667em\lower.7ex\hbox{E}\kern-.125emX}}
\begin{document}
 \title{\LARGE SkyCharge: Deploying Unmanned Aerial Vehicles for Dynamic Load Optimization in Solar Small Cell 5G Networks}
\author{\IEEEauthorblockN{}}

	\author{\IEEEauthorblockN{Daksh Dave, Vinay Chamola,~\IEEEmembership{Senior~Member,~IEEE}, Sandeep Joshi,~\IEEEmembership{Senior~Member,~IEEE}, and \\ Sherali Zeadally,~\IEEEmembership{Senior~Member,~IEEE}}\\
  \thanks{D. Dave, V. Chamola, and S. Joshi are with the Department of Electrical and Electronics Engineering, Birla Institute of Technology and Science Pilani, Pilani, Rajasthan, 333031 India. S. Zeadally is with the College of Communication and Information, University of Kentucky, Lexington, KY (e-mail: f20180391@pilani.bits-pilani.ac.in, vinay.chamola@pilani.bits-pilani.ac.in, sandeep85joshi@gmail.com, szeadally@uky.edu).}
	}
	
\maketitle
\begin{abstract}
The power requirements posed by the fifth-generation and beyond cellular networks are an important constraint in network deployment and require energy-efficient solutions. In this work, we propose a novel user load transfer approach using airborne base stations (BS) mounted on drones for reliable and secure power redistribution across the micro-grid network comprising green small cell BSs. Depending on the user density and the availability of an aerial BS, the energy requirement of a cell with an energy deficit is accommodated by migrating the aerial BS from a high-energy to a low-energy cell. The proposed hybrid drone-based framework integrates long short-term memory with unique cost functions using an evolutionary neural network for drones and BSs and efficiently manages energy and load redistribution. The proposed algorithm reduces power outages at BSs and maintains consistent throughput stability, thereby demonstrating its capability to boost the reliability and robustness of wireless communication systems.
\end{abstract}


\section{Introduction}\label{sec:Intro}
Over the past decade, there has been an exponential increase in the number of mobile users, with several billions of subscribers at present \cite{haibeh2022survey}. To meet the needs of this growing user base, the networking and communications industry is rapidly expanding, emphasizing the importance of continued investment in this sector.
Recent advances in the fifth generation (5G) and beyond communication networks have led to a rise in user-defined networking and several telecommunication systems, together with a high load on small cell BSs. All these developments have led to increased interest in the industry and academia in the field of intelligent green communication systems. As a result, solar-powered base stations (BSs) are gaining popularity because (i) they are a green solution for reducing the carbon footprint of the network operators, (ii) they can reduce operating expenditure, and (iii) they provide a means for extending cellular coverage in regions without a reliable power grid infrastructure \cite{chamola2016power}.

As of 2022, there were approximately 6.5 million BSs across the world \cite{Waring.2017}, and more than 70,000 renewable-energy-powered BSs were in operation globally \cite{baidas2022renewable}. Most of these BSs operate on fossil fuel generators, which have high operating costs and, hence, are suitable for going green. Considering the drive and need to reduce the overall carbon footprint and control global warming, we have witnessed the emergence of many green solutions that have been proposed in recent years.

Maintaining superior quality of service is vital in green cellular networks, particularly where solar power is predominant. Variations in solar energy harvest cause network instability, impacting reliability and operational longevity. This instability is exacerbated by unanticipated demand surges, causing network disruptions and substantial quality of service deterioration— a critical concern in remote or disaster-stricken areas in dire need of dependable connectivity. In economically and logistically constrained regions, the deployment and upkeep of conventional power infrastructures are challenging and costly. Current frameworks fail to effectively manage and adapt energy resources in varying operational contexts, resulting in compromised quality of service and high operational losses. These shortcomings are emphasized by traditional power grids, which experience significant energy losses and are resource-intensive \cite{chamola2015multistate}. The need for adaptable solutions such as drone-based connectivity is required to address these multifaceted challenges.

The salient contributions of this work are encapsulated as follows.
\begin{itemize}
    \item \textbf{Hybrid Framework:} Innovating a framework that synergizes drone technology with time-series and neural networks for better resource allocation in eco-friendly small-cell stations.
    \item \textbf{Unmanned aerial vehicle (UAV) Load Management:} Employing UAVs as mobile stations for optimized network coverage in energy-deficient regions, guided by cost-effective modeling.
    \item \textbf{Smart Resource Redistribution:} Fusing genetic algorithms with long short-term memory (LSTM) networks for adaptive energy and load management in high-demand zones.
    \item \textbf{Network Reliability:} Demonstrating the model's efficacy in reducing outages and enhancing throughput; pivotal for the robustness of future 5G networks.
\end{itemize}
The paper's structure unfolds as follows. Section II reviews the literature. Section III. A introduces our model, with Section III.B detailing the cost function. Section III.C discusses load balancing and optimization strategies. Section IV evaluates performance, supported by simulations. The paper concludes with Section V, highlighting the findings and implications.
\begin{figure*}[]
    \centering
    \includegraphics[width=\linewidth, frame]{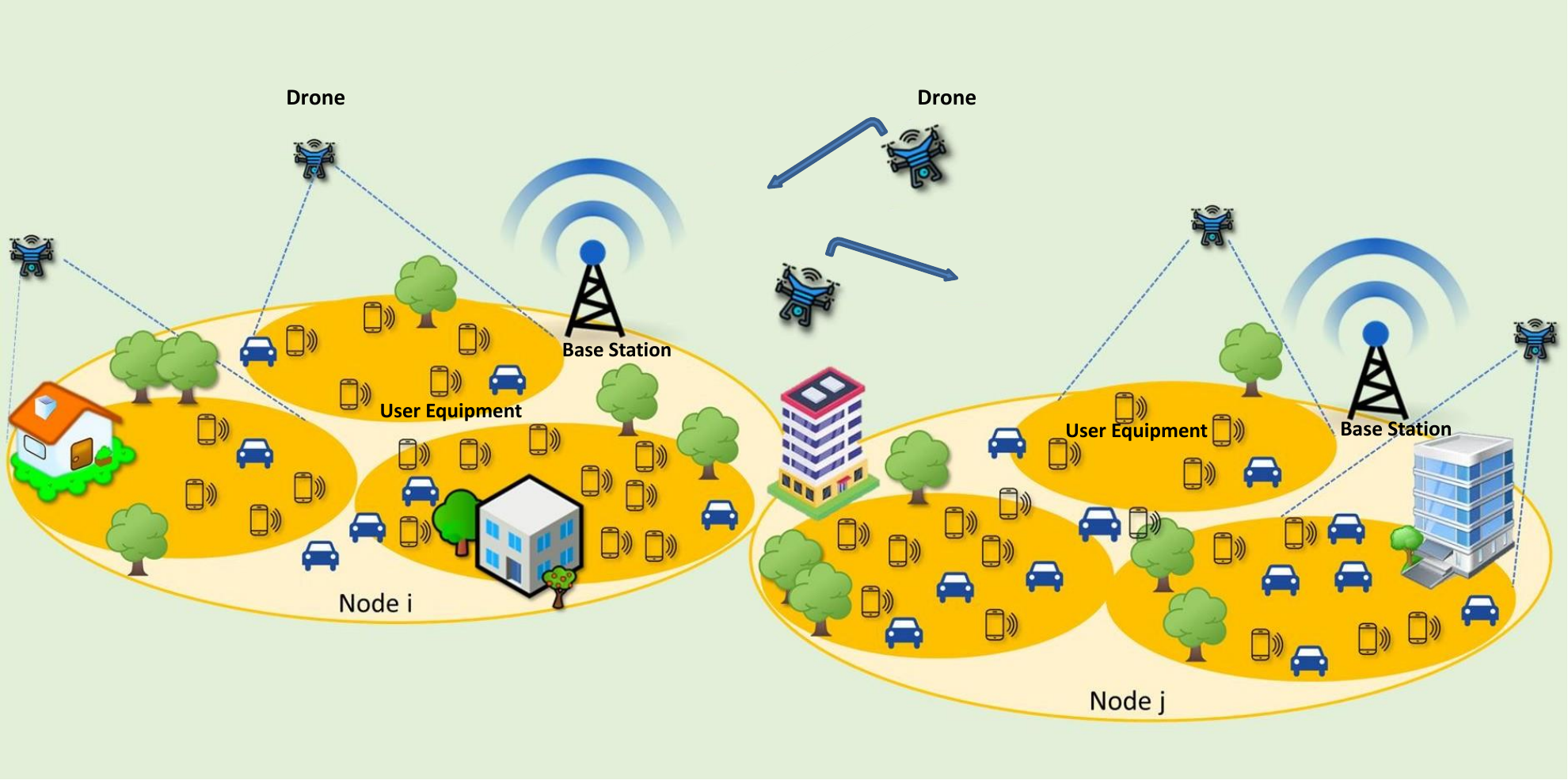}
    \caption{Representative system model depicting the small cell BS energy transfer mechanism between two nodes along with the drone exchange.}
    \label{fig:my_label_1}
\end{figure*}
\section{Related Works}\label{sec:RealtedWork}
Given the intrinsic challenge of optimizing the energy utilized from the grid and maintaining the required quality of service for efficient energy savings in solar cell small cell BSs, it is crucial to explore innovative solutions that can harmonize energy conservation with service quality. The evolving paradigm in green communication systems is paving the way for the emergence of intelligent architectures in 5G and beyond networks. In this context, the role of drones in providing a green solution is promising. The subsequent section describes existing research endeavors in small cell BSs, emphasizing intelligent drone deployments aimed at optimal positioning and power transfer efficiency determined by user mobility and various network constraints \cite{sobouti2020efficient}. In \cite{appSci2023}, the authors developed a model leveraging UAV-aided Mobile Edge Computing for load balancing and task offloading in high-demand areas, utilizing deep reinforcement learning for energy optimization. However, that work did not consider bandwidth constraints, data protection during offloading, and mobility issues between edge servers in multi-user scenarios. Extending this work, the authors of \cite{Mob_Edge} integrate differential evolution and deep reinforcement learning to maintain a balanced load and assured quality of service in IoT nodes. However, the inherent limitations of reinforcement learning, such as slower convergence rates, challenges in learning efficiency from temporal data, and scalability concerns, can impede the model's stability and applicability. Meanwhile, in \cite{2021}, the authors propose a model for drone load balancing and user equipment's data rate fairness in multi-drone networks, demonstrating enhanced stability and performance, but its reliance on static user equipment assumptions raises practicality concerns. While many studies are available, there is a noticeable absence of provisions for intelligent resource optimization and redistribution among macro BSs with small cell BSs \cite{piovesan2020joint}. The recent study by the authors of \cite{DroneBS} shows that drone-mounted aerial BSs increase terrestrial wireless network capacity and connectivity through dynamic altitude adjustments and direct line-of-sight links to the small cell BSs. These drone BSs improve the existing cellular system in terms of unprecedented mobility, flexibility, and on-demand network coverage in diverse regions, but challenges in optimal drone BS positioning and deployment in dense areas must be addressed in 5G and beyond.

\section{Methodology}\label{sec:Methodology}

\subsection{Proposed  Model}
In this paper, we focus on the incorporation of UAV to set up a network that is reliable and exhibits optimal user load re-balancing through capacity enhancement. We employ a methodology that integrates a cost-based framework with LSTM networks, aiming to accommodate higher user densities by efficiently managing the associated surge in user load and energy consumption.

We establish a model within the small cell BS, acting as a hub for the positioning of UAVs. This model plays a crucial role in identifying the optimal locations for UAV transfers on demand. It ensures network stability by diminishing a cost function tied to areas of high user demand and the deployment of UAVs. This model evaluates the network performance, aiming for comprehensive coverage in high-traffic areas. This approach seeks to lessen allocation mishaps and improve the consistency of data rate delivery.

In this construct, \( n \) UAVs are posited, each proficient in managing \( R_n \) service requests within a zone managed by a small cell BS. When service requests, \( R_s \), exceed the capacity of the small cell BS, the quality of service declines, leading to network outages. In this paper, we do not investigate the optimal spatial arrangement of drones within the small cell BS. The drones are replenished via strategically located charging stations, vital components in our model, reflecting operational and energy constraints. Our study’s mathematical construct does not consider the spatial arrangement of small cell BSs, focusing instead on energy usage, drone mobility, and service delivery. In the practical application and evolution of our model, it is critical to understand the energy dynamics and operational feasibility of utilizing drones as interim energy sources when the BSs have insufficient energy. The drones, serve as auxiliary energy conduits, designed to alleviate service interruptions and not as direct energy suppliers to the BSs, thereby focusing on optimizing the balance between energy conservation, operational mobility, and user consumption needs. The energy retention and relay capacity of the drones are mathematically modeled in our study to align with the realistic energy demands of cellular users and the inherent energy constraints of the operational environment, ensuring enhanced service reliability and a sustainable auxiliary support mechanism in energy-constrained situations.
\vspace{-0.7em}

\subsection{Mathematical Modelling and Cost Function Formulation}

To adequately support a high number of user requests at a BS, we determine the required number of UAVs, denoted as \(N_{\text{{req}}}\), using the formula \(N_{\text{{req}}} = \left\lceil \frac{R_s}{R_n} \right\rceil\). This equation ensures that sufficient UAVs are deployed to meet the high demand at each BS. Our methodology focuses on defining an appropriate cost function for individual demand zones and UAVs. This involves allocating UAVs to small-cell BSs and developing a refined cost function. The optimization of this function is then carried out through an evolutionary neural network approach, enabling the establishment of a well-balanced allocation of UAVs, which is crucial for addressing varying user demands efficiently and effectively. In the advanced model, UAVs operate at an altitude $H$, supervising an area, denoted as $\mathcal{A}$, containing $u$ users. \( R(\theta) \) represents the maximum distance that a drone can cover at a height \( H \) while maintaining good quality of service. Here, \( \theta \) is the angle between the User Equipment (UE) and the drone and is within the range \( [\theta_{\text{{min}}}, \theta_{\text{{max}}}] \) to maintain line of sight (LOS) conditions and QoS. Furthermore, \( L \) represents the length of the area covered by the SCBS(Small Cell BS).
In the study of user service requests, denoted as \(R_{\text{s}}\), with an arrival rate of \(\lambda\) and an average packet size of \(1 / \delta\), we calculate the delay or load (\(\mathcal{L}\)) experienced by a user at position \(\mathcal{Y}\) as
\begin{equation}
\mathcal{L}(\mathcal{Y}) = \frac{\lambda}{\Omega \log (1 + \text{SINR}(\mathcal{Y})) \times \delta}\,,
\end{equation}
where \(\Omega\) signifies the system bandwidth. The signal-to-interference-plus-noise Ratio (SINR) for the \(i^{th}\) UAV at location \(\mathcal{Y}\), crucial for understanding interference among UAVs within the same frequency spectrum, is given by
\begin{equation}
\text{SINR}(\mathcal{Y}) = \frac{\frac{E_{t} \kappa}{D_{i\mathcal{Y}}^{\beta}}}{\sum_{j=1, j \neq i}^{n} \frac{E_{t} \kappa}{D_{j\mathcal{Y}}^{\beta}} + {E}_{noise}}\,,
\end{equation}
where \(E_{t}\) represents the UAVs' transmission power, and \(\kappa\) is a constant that integrates geometric factors influenced by the heights of the transmitter and receiver antennas. \(D_{i\mathcal{Y}}\) marks the distance from the \(i^{th}\) UAV to the user equipment (UE) at \(\mathcal{Y}\), \(\beta\) is the path loss exponent, and \(E_{noise}\) denotes the noise power spectral density.
The spectral efficiency for a user at \(\mathcal{Y}\), under a Round Robin scheduling policy, is quantified by the effective throughput (\(E_{\text{eff}}\)), is given as
\begin{equation}
E_{\text{eff}} = \Omega \times \frac{\log_2(1 + \text{SINR}(\mathcal{Y}))}{u_{a}}\,.
\end{equation}
We assess the area load (\(\Lambda_a\)) by integrating the load across area \(\mathcal{A}\), formulated as
\begin{equation}
{\Lambda_a} = \int_{\mathcal{Y} \in \mathcal{A}} {\Lambda}(\mathcal{Y}) d\mathcal{Y}\,.
\end{equation}
To enhance our model's accuracy and efficiency, we refine the cost function by incorporating factors like capacity, latency, line of sight availability, and coverage. This refinement leads us to define \(\Phi_d\), a sophisticated density function that qualifies the concentration of users based on their request patterns. This function is influenced by the number of active users (\(u_a\)), packet loss (\(\Pi_d\)), service requests (\(R_s\)), and the cell's user capacity (\(\Theta_r\)). We differentiate between two versions of \(\Phi_d\): one for an area (\(\Phi_d^{\mathcal{A}}\)) and another for UAVs (\(\Phi_d^{\mathcal{U}}\)), as
\begin{equation}
\Phi_d^{\mathcal{A}} = \min \left( \frac{\left( \frac{u_a}{\Theta_{r}} \right)^{R_{s}} e^{-\left( \frac{u_a}{\Theta_r} \right)}}{R_{s} !} \right)\,,
\end{equation}
\begin{equation}
\Phi_d^{\mathcal{U}} = \min \left( \frac{\left( \frac{\Lambda_a}{n} \right)^{R_n} e^{-\left( \frac{\Lambda_a}{n} \right)}}{R_{n} !} \right)\,,
\end{equation}
where
\(\Phi_d^{\mathcal{A}}\) signifies the distribution of users over \(\mathcal{A}\), where higher values indicate a need for additional UAVs, and lower values suggest efficient connectivity. Conversely, \(\Phi_d^{\mathcal{U}}\) reflects the unmet user requests in \(\mathcal{A}\) relative to the number of deployed UAVs, with higher values signaling the necessity for more UAVs.
The active-to-total user ratio (\(u_a/\Theta_r\)) plays a pivotal role in these functions. To optimize the cost functions \(\Phi_d^{\mathcal{A}}\) and \(\Phi_d^{\mathcal{U}}\), they must satisfy the constraint 
\begin{equation}
\sqrt{\frac{1}{R_{s}} \sum_{i=1}^{R_{s}} \left( \Theta_r^{i} - \Pi_d^{i} \right)} \leq \frac{u_a}{\Theta_r}\,.
\end{equation}
In the contextual model, \(E_{\text{{BS}}}\) represents the energy conserved at the BS due to offloading, expressed as
\begin{equation}
E_{\text{BS}} = e_{\text{BS}} \times (L_a + E_{gen} - T_{\text{charge}})\,,
\end{equation}
where \(e_{\text{{BS}}}\) denotes the energy consumption per unit load at the BS, \(T_{\text{charge}}\) signifies the charging time of the drones in the energy consumption model, and \(L_a\) is the offloaded load. Conversely, \(E_{\text{{UAV}}}\) indicates the energy expenditure of each UAV, given as
\begin{equation}
E_{\text{UAV}} = e_{\text{UAV}} \times (d \times t + L_a - T_{\text{charge}})\,,
\end{equation}
where \(e_{\text{UAV}}\) encompasses energy consumption per unit distance, per unit time, and per unit load, with \(d\), \(t\), and \(L_a\) representing the distance flown, time spent servicing, and the load being served, respectively.
The constants \(\delta\), \(\beta\), \(\gamma\), and \(\epsilon\) are introduced to balance the significance of various energy aspects in the overall costs. The energy consumed during drone movement considering UAV trajectory optimization is given as
\begin{equation}
E_{\text{travel}} = e_{\text{travel}} \times (d \cdot T_{\text{mobility}})
\end{equation}
where \(T_{\text{mobility}}\) denotes the time drones spend moving from one area to another.
The energy consumption due to node communications, taking into account various channel conditions, is integrated as
\begin{equation}
E_{\text{comm}} = \eta \sum_{k=1}^{K} E_t \log_2\left(1 + \frac{G_k E_t}{E_{noise}}\right)\,,
\end{equation}
where \(\eta\) is the energy consumption coefficient, \(E_t\) is the transmit power, \(G_k\) is the channel gain, and \(E_{noise}\) is the noise power for the \(k^{th}\) communication link, summing over all \(K\) communication links.
For optimal performance, it is necessary to minimize cost-function constraints representing users with unresolved service requests. Thus, we refined our energy-aware and mobility-conscious cost functions for each area and UAV. Intermediate variables representing combined energy terms are given as
\begin{equation}
E_{\text{{total}}}^{\mathcal{A}} = - \delta E_{\text{{BS}}} + \gamma E_{\text{{travel}}} + \epsilon E_{\text{{comm}}},
\end{equation}
\begin{equation}
E_{\text{{total}}}^{\mathcal{U}} = \beta E_{\text{{UAV}}} + \gamma E_{\text{{travel}}} + \epsilon E_{\text{{comm}}}\,.
\end{equation}
Consequently, the refined cost function is given as
\begin{equation}
C_{\phi}^{\mathcal{A}} = \min \left( a_i \Phi_d^{\mathcal{A}} \Lambda_a \left( \zeta_{1} R_{s} + \zeta_{2} \Theta_r + E_{\text{{total}}}^{\mathcal{A}} \right) \right)\,,
\end{equation}
for scenarios where line of sight is not taken into account, and is given as
\begin{equation}
C_{\phi}^{\mathcal{U}} = \min \left( a_i \Phi_d^{\mathcal{U}} D_{i\mathcal{Y}}^{\beta} \left( \zeta_{1} R_{s} + \zeta_{2} u_a + E_{\text{{total}}}^{\mathcal{U}} \right) \right)\,,
\end{equation}
when line-of-sight considerations are taken into account.
The availability parameter, \(a_i\), represents the real-world scenarios of drone availability, where \(a_i = 1\) signifies that drone \(i\) is available and \(a_i = 0\) indicates its unavailability. This parameter, integrated with \(E_{\text{travel}}\) and \(E_{\text{comm}}\), enriches the model by accounting for mobility and communication energy consumption under various channel conditions, thereby enhancing the adaptability and performance analysis of the network with respect to drone availability. The consideration of line of sight is crucial for ensuring uninterrupted connectivity.
Our methodology employs a Long Short-Term Memory (LSTM) model to forecast the power expenditure (\(P_{\text{{LSTM}}}\)) for each BS independently, based on user density sequences and the corresponding energy expenditure matrix (\(E_i\)) for every BS. This enables the model to discern the underlying patterns and dependencies within the data for each station. Consequently, for each BS \(j\), a unique \(P_{\text{{LSTM}},j}\) is computed, facilitating the integration of these values into our comprehensive optimization framework. The overall cost optimization equation is thereby modified to incorporate \(P_{\text{{LSTM}},j}\) alongside the cost associated with UAV communication with the backend or server (\(C_{\text{backend}}\)) as 
\begin{scriptsize}
\begin{equation}
C_{\phi}^{\mathcal{O}} = \min \left( \frac{1}{n} \sum_{i=1}^{n} \left( C_{\phi}^{\mathcal{U}} + C_{\text{backend}} \right)_{i} + \sum_{j=1}^{\mathcal{A}_T} \left( \frac{C_{\phi}^{\mathcal{A}} + P_{\text{LSTM},j}}{\mathcal{U}_T} \right)_{j} \right),
\end{equation}
\end{scriptsize}
where \(j\) indexes the BSs, and \(P_{\text{{LSTM}},j}\) denotes the predicted power expenditure by the LSTM model for the \(j^{th}\) BS. This approach provides a detailed representation of the energy dynamics across various BSs, leading to a more robust and comprehensive optimization solution.

In the refined model, \(\mathcal{U}_T\) signifies the number of UAVs allocated to a specific area, and \(\mathcal{A}_T\) denotes the total number of demand areas. The weighting factor \(\lambda\) is meticulously determined based on factors such as historical user demand, variability in service requests, and expected network load, which in turn influences the predictive term \(P_{\text{{LSTM}}}\)'s impact on the overall cost. The deployment of additional UAVs offers supplementary resources, enhancing transmission power optimization, which leads to increased throughput and reduced latency. This holistic approach ensures that energy considerations are seamlessly integrated into the operational framework, providing a comprehensive outlook on network connectivity, user service, and energy efficiency.

\titlespacing*{\subsection}{0pt}{\baselineskip}{\baselineskip}
\vspace{-0.25em} 
\begin{algorithm}
\caption{Evolutionary UAV neural network training}
\label{algorithm1}
\begin{algorithmic}[1]
\State \textbf{Input Data:} train, test, valid, epoch, learning\_rate, $U$, population\_size, generations
\State \textbf{Inputs:} $u_a$, $\Theta_r$, $\Lambda_a$, $n$, $R_s$, $\zeta_1$, $\zeta_2$, $E_{\text{total}}^{\mathcal{A}}$, $E_{\text{total}}^{\mathcal{U}}$, $D_{i\mathcal{Y}}^{\beta}$

\State Initialize population of neural networks
\For{g = 1 to generations}
    \For{each individual in population}
        \For{i = 1 to epoch}
            \For{each batch in train}
                \State Compute $P_{\text{{LSTM}},j}$
                \State Compute $\phi_d^{\mathcal{A}}$, $\phi_d^{\mathcal{U}}$, $C_{\phi}^{\mathcal{A}}$, $C_{\phi}^{\mathcal{U}}$, $C_{\phi}^{\mathcal{O}}$                 \State Update $U$ to minimize $C_{\phi}^{\mathcal{O}}$
                \State loss = $\overline{C_{\phi, pred}^{\mathcal{O}}}$ - $C_{\phi, true}^{\mathcal{O}}$
                \If{constraint violated}
                    \State loss = loss + $\lambda \cdot \text{penalty}$
                \EndIf
            \EndFor
            \State Evaluate model on valid set
            \If{optimality criterion met}
                \State \textbf{break}
            \EndIf
            \State Adjust hyperparameters
        \EndFor
    \EndFor
    \State Perform selection, crossover, mutation and replacement on population
    \If{stopping criterion met}
        \State \textbf{break}
    \EndIf
\EndFor
\State Evaluate the best individual on test
\If{criterion met and $U$ available}
    \State Fine-tune best model
\EndIf
\State \textbf{Return} trained\_model
\end{algorithmic}
\end{algorithm}
\begin{figure*}[!ht]
    \centering
    \includegraphics[width=\textwidth, height=3.7in]{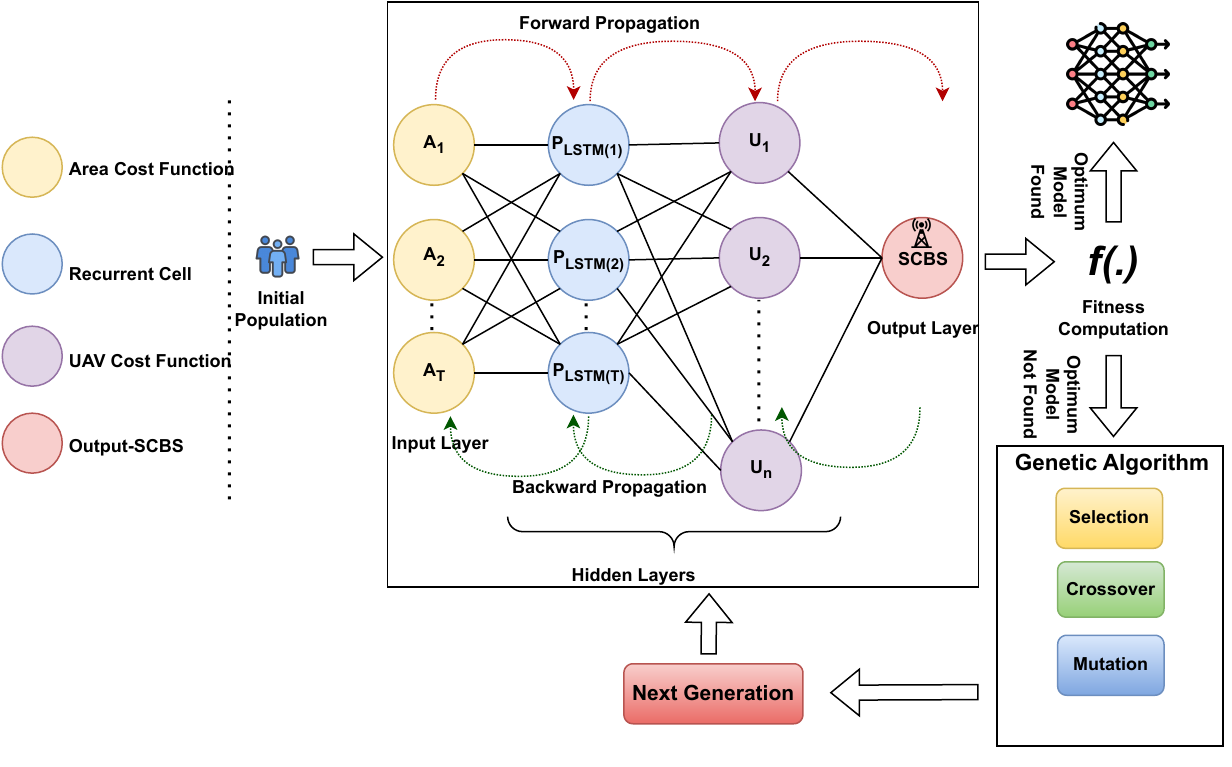}
    \caption{Model architecture.}
    \label{fig:modelArchitecture}
\end{figure*}
\subsection{Efficient Load Balancing and Cost Optimization}
\vspace{-0.8em}
Fig. \ref{fig:modelArchitecture} shows a hybrid model SkyCharge: An evolutionary Neural Network-based LSTM model designed to handle complex time-series data, leveraging global optimization techniques of Evolutionary Strategies to avoid local optima prevalent in Reinforcement Learning approaches. Evolutionary Strategy's methodical global search complements LSTM's swift convergence, ensuring robust solutions in non-convex optimization landscapes, crucial for real-time, scalable network optimizations. We use genetic algorithms in our model to explore and optimize the solution space for UAV allocation, allowing for the identification of optimal or near-optimal solutions for minimizing costs, even in the presence of complex, nonlinear relationships and constraints. Additionally, the genetic algorithm's inherent parallelism and adaptability make it an excellent choice for dynamically adjusting the model to varying demand intensities and distributions in real-world scenarios. This hybrid model enables adaptive refinement of network parameters and architectures, iterating and evolving solutions based on their fitness, which is quantified by their ability to minimize the associated cost functions and balance loads across various BSs. This integration ensures real-time resource allocation and energy balancing.

Our model optimizes the quantity and deployment of UAVs to energy-depleted BSs, aiming to reduce both UAV-related and overall network costs. The cost functions are utilized to minimize total energy use, operational costs, and resource inefficiencies. Our model optimizes demand by redistributing drones to small cell BSs projected to experience high energy demand, thereby improving the overall network efficiency. In our model, the output layer is represented by the demand areas of small cell BSs, while the hidden layers are determined, and dependencies within user demand trends and BS energy consumption patterns incorporating both LSTMs and UAVs to capture complex patterns and dependencies within the user demand trends and BS energy consumption patterns. The LSTM layers are instrumental in capturing temporal dependencies and facilitating the prediction of energy deficits in real-time, thereby enabling the model to preemptively allocate resources to meet the projected demand. UAVs play a crucial role in ensuring the targeted delivery of energy to the areas most in need, effectively optimizing the energy distribution within the network. The ultimate goal of our model is to minimize the cost functions associated with each BS, facilitating a balanced load across the network. This is achieved through strategic rearrangements of neural patterns to form a stable and optimized network with minimized cost functions, allowing for seamless adaptation to varying energy demand intensities. 

Algorithm-\ref{algorithm1} presents the refined methodology and describes the steps of the evolutionary UAV neural network Training procedure combined with \(P_{LSTM}\). The process starts by including multiple types of input such as the number of active users-\(u_a\), the maximum number of users the small cell BSs can handle-\(\Theta_r\), load density-\(\Lambda_a\), and so on. The algorithm begins with the initialization of a population of neural networks that undergo numerous generations, each consisting of multiple individuals. The evolutionary neural network is configured to run for a maximum of 100 epochs, with early stopping implemented if the validation loss does not improve for 10 consecutive epochs. The computational cost for training the neural network is moderate due to the complexity of the model and the size of real-world datasets. Training is estimated to require approximately 4 hours on a system with a 16-core CPU and a GPU with 24 GB of VRAM. Given the dynamic nature of network traffic, the model is designed for retraining on a quarterly basis scheduled using a CRON job to adapt to new patterns, or as needed when a certain threshold of change in network data is detected which can again be configured through a CRON job.

The initial population of neural networks is created by sampling a diverse set of hyperparameters within predefined ranges. These ranges are determined based on the network's architectural constraints and prior empirical benchmarks. For instance, the learning rate is varied between 0.001 and 0.1, the number of hidden layers can range from 1 to 5, and the number of neurons per layer might be set between 50 and 500. Additional hyperparameters include the type of activation functions (e.g., ReLU, sigmoid, tanh) and the dropout rate, which is typically varied from 0 to 0.5 to prevent overfitting. The initialization also considers LSTM-specific parameters such as the number of LSTM units, ranging from 20 to 200, and the forget gate bias, selected between 1 and 5. This approach ensures a comprehensive search across the architectural space, providing a robust starting point for the evolutionary optimization process.
 Our population size is set to 50 individuals per generation, a figure optimized for computational efficiency and sufficient genetic diversity. We use a batch size of 128 for training, which balances the gradient estimation accuracy and memory constraints. In each generation, every individual undergoes a series of training phases. Here, the pLSTM computes \(P_{\text{{LSTM}},j}\) which integrates with the computation of pivotal elements such as area, UAV, and overall cost functions— \(C_{\phi}^{\mathcal{A}}\), \(C_{\phi}^{\mathcal{U}}\), and \(C_{\phi}^{\mathcal{O}}\), respectively. Concurrently, \(U\) is optimized to balance \(C_{\phi}^{\mathcal{A}}\) and \(C_{\phi}^{\mathcal{U}}\), distributing the overall network load. Any violation of predetermined constraints incurs a penalty during the loss computation phase.

After every epoch, the algorithm evaluates the different potential solutions in the genetic algorithm population using the validation data, subsequently fine-tuning the hyperparameters to enhance the model's efficacy. It includes checks for constraints such as UAV energy limits, payload capacities, and communication bandwidth post-batch, applying penalties to the loss function for any violations to ensure that solutions stay within practical operational limits. This iterative process continues until an optimality criterion, which could be a predefined performance threshold or a maximum number of epochs, is met. Genetic operators like selection, crossover, mutation, and replacement are then applied to the population after each generation, iterating until a stopping criterion is reached. Upon completing the generational cycles, the algorithm evaluates the elite individual using the test data, and further refinement is conducted if necessary and if drones (\(U\)) are available, based on predefined criteria. Finally, the algorithm outputs the optimally trained model, which orchestrates the strategic redistribution of UAVs to high-demand areas, iteratively minimizing the associated cost functions.

\section{Performance evaluation and simulation Results}
\begin{table}[h!]
\centering
\footnotesize
\begin{tabular}{|l|r|l|}
\hline
\textbf{Variable} & \textbf{Value} & \textbf{Short Description} \\
\hline
\( n \) & 10 & Total number of UAVs posited \\
\( \mathcal{A}_T \) & 5 & Total energy demand areas \\
\( R_n \) & 50 & Number of UAV service requests \\
\( R_s \) & 100-150 & Number of user service requests \\
\( H \) & 150m-450m & UAV altitude \\
\( \mathcal{A} \) & 25km\textsuperscript{2} & Total area by SCBS \\
\( u \) & 1000 & Number of users \\
\( \lambda \) & 2 req/s & Arrival rate of requests \\
\( \frac{1}{\delta} \) & 200 Bytes & Mean packet size \\
\( \frac{\lambda}{\delta} \) & 0.01 req/s & Offered traffic \\
\( \Omega \) & 20 MHz & System bandwidth \\
\( E_t \) & 10 W & UAV transmission power \\
\( \kappa \) & 1.5 & Geometrical aspects \\
\( \beta \) & 3 & Path loss exponent \\
\( E_{\text{noise}} \) & $-174$ dBm/Hz & Noise power density \\
\( u_a \) & 200 & Number of active users \\
\( \Pi_d \) & 5\% & Packet loss or call drops \\
\( \Theta_r \) & 300 & Cell user capacity \\
\( L \) & 500m & BS length \\
\( R(\Theta) \) & 100m & Maximum UAV radius \\
\( \eta \) & 0.5 & Energy consumption coefficient \\
\( G_k \) & 10 dB & Channel gain \\
\( e_{UAV}, v \) & 0.02 J/m, 20m/s & UAV energy/unit distance, drone speed \\
\hline
\end{tabular}
\caption{Summary of simulation parameters}
\label{tab:simulation_parameters}
\end{table}

\begin{figure*}[ht!]
    \centering
    \begin{subfigure}[t]{0.45\textwidth}
        \centering
        \includegraphics[width=\textwidth]{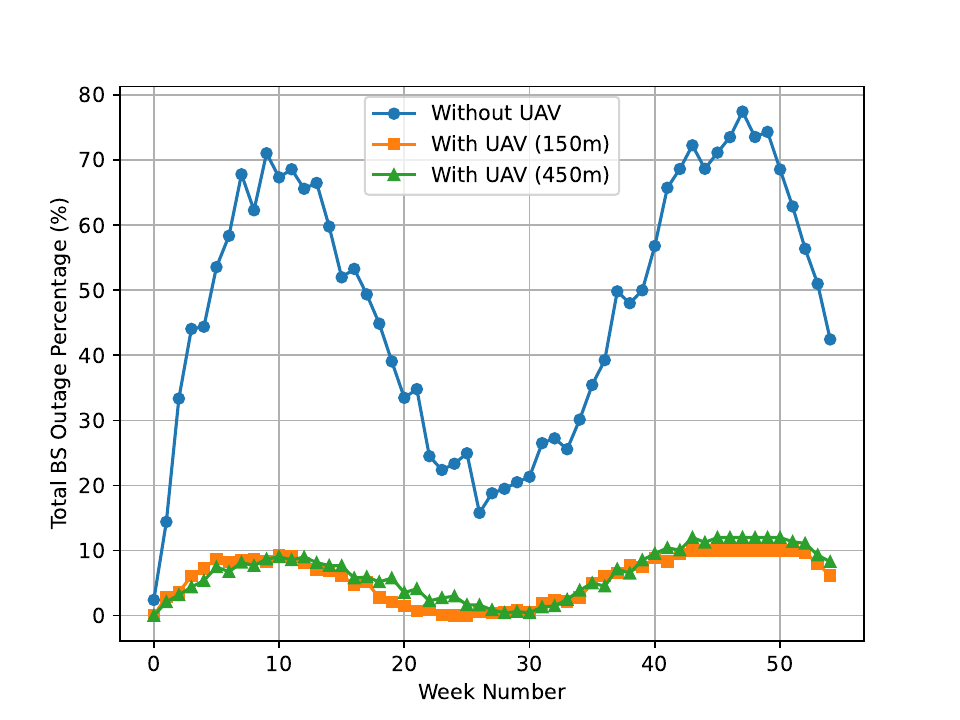}
        \caption{BS power outages versus Week Number.}
        \label{fig:outageVsWeek}
    \end{subfigure}
    \begin{subfigure}[t]{0.45\textwidth}
        \centering
        \includegraphics[width=\textwidth]{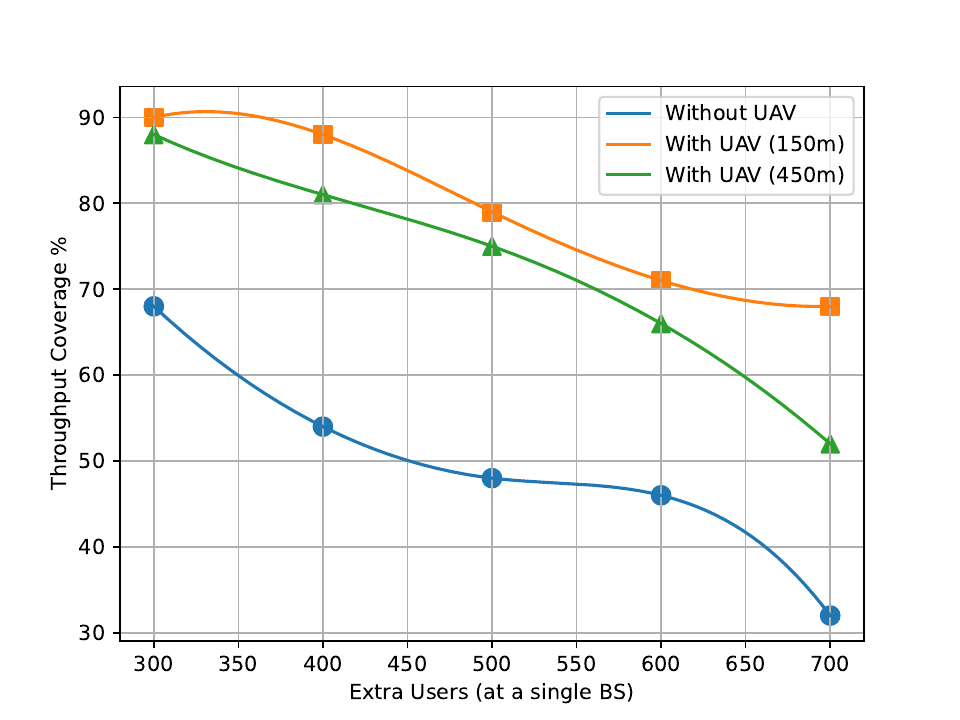}
        \caption{Throughput Coverage versus Extra Users.}
        \label{fig:throughputVsUsers}
    \end{subfigure}
    \caption{Analysis of BS power outages and throughput coverage under different scenarios.}
    \label{fig:comparativeAnalysis}
\end{figure*}

\begin{figure*}[ht!]
    \centering
    \begin{subfigure}[t]{0.45\textwidth}
        \centering
        \includegraphics[width=\textwidth]{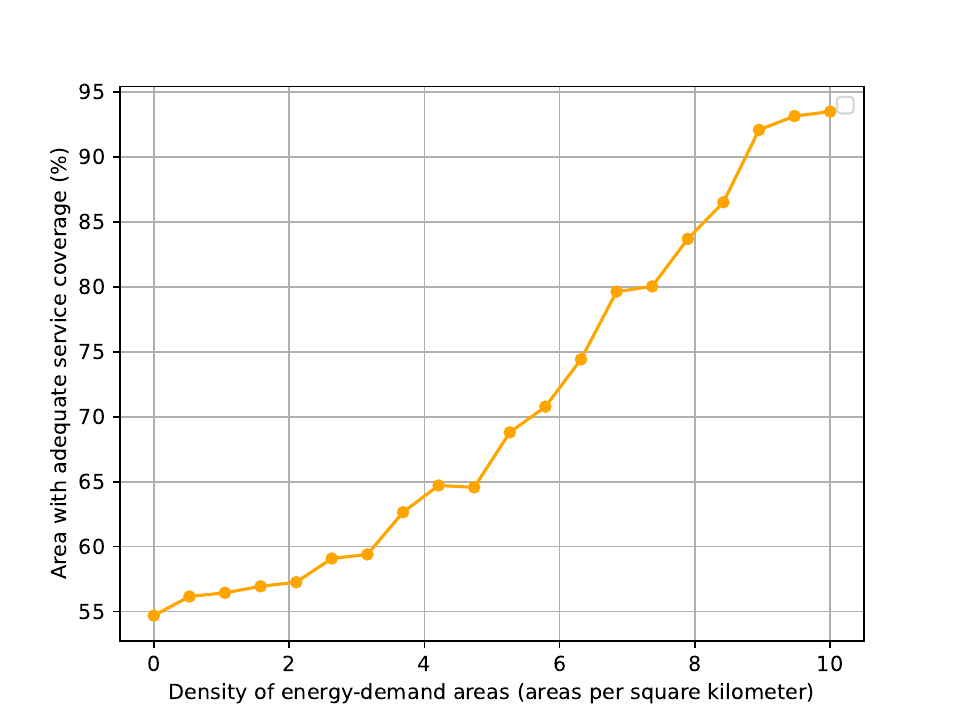}
        \caption{Service Coverage Area versus Energy-Demand Area Density.}
        \label{fig:coverageVsUserDensity}
    \end{subfigure}
    \begin{subfigure}[t]{0.45\textwidth}
        \centering
        \includegraphics[width=\textwidth]{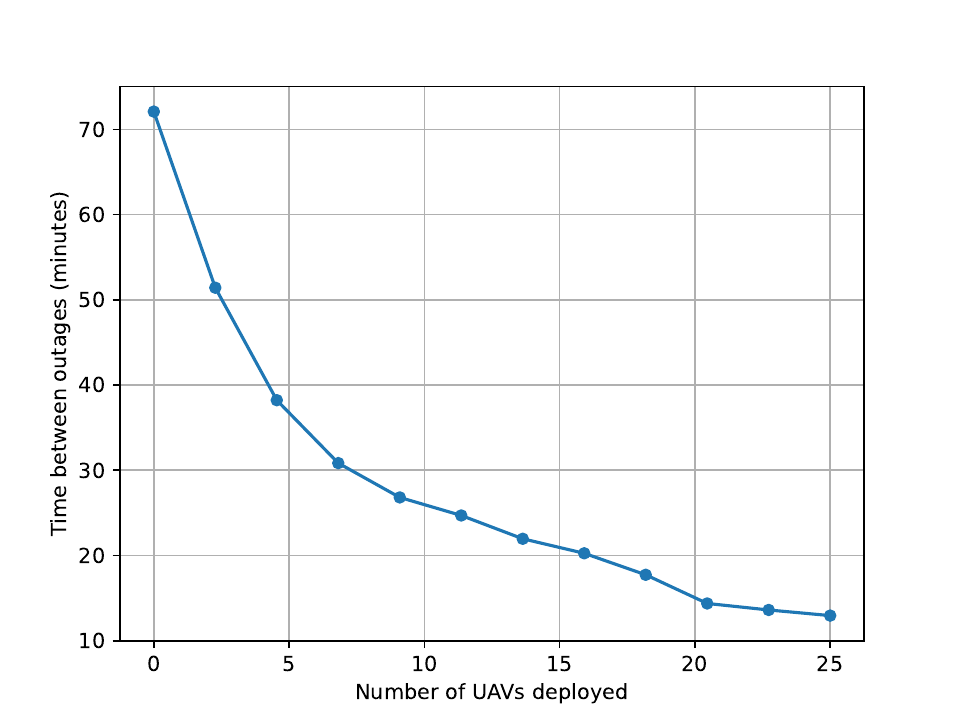}
        \caption{Time Between BS Outages versus Number of UAVs.}
        \label{fig:outagetimevsNoUAVs}
    \end{subfigure}
    \caption{Analysis of service area coverage and time between BS outages under different scenarios.}
    \label{fig:comparativeAnalysis2}
\end{figure*}

\begin{table}[h!]
\centering
\begin{tabular}{|l|c|c|c|}
\hline
Metric & RMSE & MAE & \( R^2 \) \\
\hline
Value & 12.34 & 9.45 & 0.76 \\
\hline
\end{tabular}
\caption{Updated evaluation metrics of the LSTM model}
\label{tab:updated_evaluation_metrics}
\end{table}

The dataset \cite{gorla2022decentralized} used comprises hour-wise solar data from five base stations (BSs) in Jaipur, Rajasthan, India. The dataset is structured in a 24-hour and 365-day notation. We conducted simulations utilizing Google Colab Pro and Matlab R2021a, with the application of data preprocessing techniques for optimal analysis. The refined dataset yielded matrices indicating user density and energy metrics at each BS every hour. Table -\ref{tab:simulation_parameters} presents the simulation parameters used in our model’s simulation.  The selection of simulation parameters within our SkyCharge framework was carefully curated, reflecting a balance between real-world operational constraints and the need for robust network performance in simulated environments. The UAV transmission power is set at 10 W, a parameter consistent with the transmission capabilities of commercial UAVs designed for communications support, as indicated by recent field trials and industry standards \cite{Zeng2016}. The packet loss rate, assumed at 5\%, is based on typical urban communication conditions and corroborated by benchmark studies in UAV-assisted cellular networks \cite{Gupta2015}. Altitude ranges, service request rates, and UAV speeds are derived from a composite of regulatory limitations, aeronautical performance data, and operational UAV network studies \cite{DGCA2021,Mozaffari2019}. These parameters are not only representative of current UAV network deployments but are also grounded in a body of literature that validates their applicability and efficacy. For instance, altitude ranges are chosen in line with DGCA regulations and existing literature on optimal UAV operational ceilings for maximum coverage with minimal interference \cite{AlHourani2014}. The service request rate is benchmarked against typical data traffic patterns observed in metropolitan 5G networks \cite{Andrews2014}, while the UAV speeds reflect operational safety and efficiency metrics \cite{Shakeri2019}. It is important to note that these parameters, while representative at the time of writing, will be subject to continuous review and adjustment in line with technological advancements and evolving regulatory frameworks.

We employed several key metrics in our study:
\begin{itemize}
  \item \textbf{Root Mean Square Error (RMSE)}: Measures the average squared differences between predicted and actual values, emphasizing larger errors.
  
  \item \textbf{Mean Absolute Error (MAE)}: Calculates the average absolute discrepancies between predicted and actual figures, displaying error magnitude.
  
\item \textbf{Coefficient of Determination ($R^2$)}: Quantifies the proportion of the variance in the dependent variable that is predictable from the independent variables, serving as an indicator of the model's goodness of fit relative to a baseline model.
  
  \item \textbf{Power Outages}: Denotes the instances of complete power loss at BSs, critical for evaluating network reliability and service continuity.
  
  \item \textbf{Throughput Coverage}: Represents the network's capacity to sustain specified data transfer speeds across an area. It is defined as the ratio of users maintaining a Signal-to-Interference-plus-Noise Ratio (SINR) above the threshold of $0.045 \, \text{bps/Hz}$.
\end{itemize}

Next, we present the results from the simulation of the modeled system.  We evaluated the LSTM model's predictive performance on energy expenditure through Table-\ref{tab:updated_evaluation_metrics} which shows that our LSTM model achieved an RMSE of 12.34, MAE of 9.45, and $R^2$ of 0.76, indicating substantial predictive capability but room for reducing larger errors.UAVs are deployed to alleviate the energy deficits of BSs, with a focus on assessing outages, network capacity, reliability, and the value of the cost function, both with and without the use of UAVs. The UAV can only establish a link with the end-user when it achieves line of sight. A significant challenge is determining the appropriate flying altitudes to avoid collisions and interferences when UAVs share altitudes, and maintaining quality of service at higher altitudes where interference is minimal. The altitudes are varied between $150 \mathrm{ft}$ and $450 \mathrm{ft}$, with improved \textit{MIMO} multi-antenna relay.

Subfigure-\ref{fig:outageVsWeek} depicts the BS power outage percentages over different weeks, showing significantly reduced outages when UAVs are deployed. The UAV deployment reduces power outages by $89.2 \%$ across all BSs compared to a small-cell network. Throughput coverage, represented in Fig.-\ref{fig:throughputVsUsers} captures the relationship between throughput coverage percentages and additional users. We observe a decrease in throughput coverage when the number of extra users increases, particularly without UAV support, dropping to 32\% with 700 additional users. The integration of UAVs at $150 \, \mathrm{m}$ and $450 \, \mathrm{m}$, however, significantly mitigates this decrease, maintaining coverage above 68\% and 52\%, respectively. This variance in throughput robustness under different UAV altitudes emphasizes the essential role of optimal UAV deployment in enhancing network resilience and maintaining throughput efficiency under increased user demands. 

Our findings, illustrated in Figure-\ref{fig:coverageVsUserDensity}, show a positive correlation between the density of energy-demand areas and service coverage enhanced by UAV deployment. This efficacy peaks in medium-density regions and plateaus in high-density regions, suggesting a limit to the benefits of increased density. Figure-\ref{fig:outagetimevsNoUAVs} further reveals that deploying UAVs reduces the time between BS outages, with diminishing returns as more UAVs are added.

The deployment of UAVs bolsters service coverage and network resilience, validating our hypothesis that UAVs are a potent solution for on-demand energy supply and outage reduction, thereby enhancing network quality. Our model outperforms traditional methods by improving coverage and reducing outages.

\section{Conclusion}
We proposed a hybrid drone-based framework leveraging UAVs, LSTM models, and genetic algorithms for energy and load management in 5G networks. This model enhances energy efficiency and network stability, showing promise for integrating renewable energy into UAVs and supporting internet of things ecosystems and ultra-reliable low-latency communication protocols. This will include exploring adaptive network topologies and utilizing quantum computing for optimization, advancing the development of efficient, self-organizing networks for 5G and beyond.

\begin{center}
    \bibliographystyle{IEEEtran}
    \bibliography{ref}
\end{center}
\end{document}